\documentclass[conference]{IEEEtran}

\usepackage{cite}

\ifCLASSINFOpdf
   \usepackage[pdftex]{graphicx}
\else
\fi
\usepackage[cmex10]{amsmath}
\usepackage{algorithmic}
\usepackage[tight,footnotesize]{subfigure}
\usepackage{color}
\usepackage{enumerate}
\usepackage{algorithm}
\usepackage{footnote}
\usepackage{url}
\begin{document}
%
\title{Adaptive Hierarchical Data Aggregation using Compressive Sensing (A-HDACS) for Non-smooth Data Field}

\author{\IEEEauthorblockN{Xi Xu}
\IEEEauthorblockA{Department of Electrical and\\Computer Engineering\\
University of Illinois at Chicago\\
Chicago,Illinois,60607\\
Email:xxu25@uic.edu}
\and
\IEEEauthorblockN{Rashid Ansari}
\IEEEauthorblockA{Department of Electrical and\\Computer Engineering\\
University of Illinois at Chicago\\
Chicago,Illinois,60607\\
Email: ransari@uic.edu}
\and
\IEEEauthorblockN{Ashfaq Khokhar}
\IEEEauthorblockA{Department of Electrical and\\Computer Engineering\\
University of Illinois at Chicago\\
Chicago,Illinois,60607\\
Email:ashfaq@uic.edu}}

\maketitle

\begin{abstract}
Compressive Sensing (CS) has been applied successfully in a wide variety of 
applications in recent years, including photography, shortwave infrared cameras,
optical system research, facial recognition, MRI, etc. In wireless sensor networks (WSNs), 
significant research work has been pursued to investigate the use of CS to 
reduce the amount of data communicated, particularly in data aggregation applications 
and thereby improving energy efficiency. 
However, most of the previous work in WSN has used CS under the assumption that data 
field is smooth with negligible white Gaussian noise. In these schemes signal 
sparsity is estimated globally based on the entire data field, which is then used to determine the CS parameters.
In more realistic scenarios, where data field may have regional fluctuations or it is 
piecewise smooth, existing  CS based data aggregation schemes yield poor compression efficiency.   
In order to take full advantage of CS in WSNs,
we propose an Adaptive Hierarchical Data Aggregation using Compressive Sensing
(A-HDACS) scheme. The proposed schemes dynamically chooses sparsity values based on signal variations
in local regions. We prove that A-HDACS enables more sensor nodes to employ CS compared 
to the schemes that do not adapt to the changing field. 
The simulation results also demonstrate the improvement in energy efficiency as well 
as accurate signal recovery. 

\end{abstract}

\IEEEpeerreviewmaketitle

\section{Introduction}
Energy efficiency is a major target in the design of wireless sensor networks due to limited 
battery power of the sensor nodes. Also, at times it is difficult to replenish battery power 
depending on the application area. Since data communication is the most basic but high energy consuming task in 
sensor networks, a plethora of research work has been done to improve its 
energy consumption~\cite{DataAgg} \cite{DataAgg1} \cite{DataAgg2} \cite{DataAgg3}.
Compressive Sensing (CS) \cite{CS} \cite{CS_Richard} has emerged as a promising technique to reduce the amount of data communicated in WSNs.  
It has been also applied in other application areas such as photography, 
shortwave infrared cameras, optical system research, facial recognition, MRI, etc.~\cite{CS_App}.
Luo et. al.~\cite{BalCS} proposed the use of CS random measurements to replace raw data 
communication in data aggregation tasks in WSNs, thus reducing the amount of data transmitted. 
However, their technique introduced redundant data communication in nodes that were farther away 
from the root node of the data aggregation tree.
Xiang et. al.\cite{HybridCS1} \cite{HybridCS2} optimized this scheme 
by reducing the data transmission redundancy.  In our previous work, We further improved CS based data aggregation by proposing a 
Hierarchical Data Aggregation using Compressive Sensing (HDACS) \cite{HDACS}  
that introduced a hierarchy of clusters into CS data aggregation model and achieved 
significant energy efficiency. 

However, most of the previous work has used CS under the assumption that data 
field is smooth with negligible white Gaussian noise. In these schemes, signal 
sparsity is calculated globally based on the entire data field. 
In more realistic scenarios, where data field may have regional fluctuations or it is 
piecewise smooth, existing  CS based data aggregation schemes will yield poor compression efficiency.   
The sparsity constant $K$ is usually a big number, with large probability, 
when the field consists of bursts or bumps. 
In such cases, the number of CS measurements $M=K \log N$ is bigger than $N$, 
where $N$ is local cluster size. In order to take full advantage of CS for its great compression capability,
we propose an Adaptive Hierarchical Data Aggregation using Compressive Sensing
(A-HDACS) scheme.The proposed schemes adaptively chooses sparsity values based on signal variations
in local regions. 
 
Our solution is based on the observation that the number of CS random measurements 
from any region (spatial or temporal) should correspond to the 
local sparsity of the data field, instead of global sparsity. 
Intuitively, it should work well because the nodes are 
more correlated with each other in a local area than the entire global area.  
Also,  it is easy to compute the local sparsity, particularly when a data aggregation scheme is 
based on a hierarchical clustering scheme.  Also, in order to compute global sparsity, 
apriori knowledge of the data field is required.  
We show that the proposed A-HDACS scheme 
enables more sensor nodes to utilize compressive sensing compared to the HDACS 
scheme~\cite{HDACS}
that employs global sparsity based compressive sensing. Using the SIDnet-SWANS \cite{SIDnet} 
sensor simulation platform for our experiments, we demonstrate 
the effectiveness of the proposed scheme 
for different types of data fields and network sizes. 
For the smooth data field with multiple Gaussian bumps, A-HDACS 
reduces energy consumption by $\approx 6\% $ to $10\%$, depending on the network size. Similarly, 
for the piecewise smooth data 
field, it reduces energy consumption by $\approx 23.36\% $ to $30.17\%$ depending on the network size.
We observe higher gains in larger network sizes. 
The experimental results are consistent with our theoretical analysis. 

The rest of paper is organized as the follows: Section II gives an overview of the existing 
CS based data aggregation schemes. In Section III, the details of the proposed A-HDACS scheme 
are presented. The analysis of the data field sparsity and its effect on CS in both HDACS and 
A-HDACS is given in Section IV. Section V shows the simulation evaluation. 

\section{Related Work}
Any conventional data collection scheme that does not involve pre-processing of data usually 
employs $O(N^2)$ data transmissions
in an $N-$node routing path. Lou et al. \cite{BalCS} were the first to examine the use of 
Compressive Sensing (CS)~\cite{CS} \cite{CS_Richard} in data gathering applications  
for large scale WSNs. Their scheme reduced the 
required number of transmissions to $O(NM)$, where $M << N$. 
According to CS \cite{CS},  $M=K \log{N}$ and $K$ is the 
signal sparsity, representing the number of nonzero entries of the signal. 
We refer to this scheme as the plain CS aggregation scheme (PCS). 
PCS requires all sensors to collectively provide to the sink
the same amount of random measurements, i.e. $M$, regardless of their location in the network. 
Note that when PCS is applied in a large scale network, $M$ may still be a large number. 
Moreover, in the initial data aggregation phase in \cite{BalCS}, nodes placed on or closer to the leaves of aggregation tree 
also transmit $M$ measurements, which is in excess of their single readings and therefore introduces redundancy in 
data aggregation.  The hybrid CS (HCS) aggregation \cite{HybridCS1}\cite{HybridCS2} 
eliminated the data aggregation redundancy in the initial phase by combining conventional data aggregation with PCS. 
It optimizes the data aggregation cost by setting a global threshold $M$ and applying CS at only those nodes 
where the number of accumulated data samples equals to, or exceeds $M$; otherwise all other nodes communicate just raw data. 
The major drawback of HCS is that only a small fraction of the sensors are able to utilize the advantage of CS scheme, 
and the required amount of data that need to be transmitted for even these nodes is still large.
Thus, an energy-efficient technique: Hierarchical Data Aggregation using Compressive Sensing (HDACS) 
\cite{HDACS} was presented based on a multi-resolution hierarchical clustering architecture and HCS. 
The central idea was to configure sensor nodes so that instead of one sink node being targeted by all sensors, several nodes, 
arranged in a way to yield a hierarchy of clusters, are designated for the intermediate data collection. 
The amount of data transmitted by each sensor is determined based on the local cluster size 
at different levels of the hierarchy rather than the entire network, which, therefore, leads to reduction
in the data transmitted, with an upper bound of $O(K\log{N})$. In other words, in HDACS the value of $N$ is different for 
different nodes. But HDACS has its own limitation. It can only solve the data aggregation problem when the data field is globally smooth with 
negligible variations, since its data field sparsity is assumed as a single constant $K$ derived from the whole data field. 
It is more desirable that we can consider more realistic scenarios when the data 
field is not relatively flat, i.e. sparsity of the data field is different for different regions of the network. 
In this work, our attention will mainly focus on 
how the fluctuations of the data field affects HDACS and we propose Adaptive HDACS (A-HDACS) to solve this problem.

\section{Proposed Adaptive HDACS (A-HDACS) Scheme}
The basic idea behind A-HDACS is that CS random measurements for 
each sensor that need to be communicated are determined by the sparsity of data field 
within each clusters at different levels of the data aggregation tree.

For consistency, we adopt the same notations as in \cite{HDACS},  showed in Table \ref{ParaDef1}. 

\begin{table}[!t]
\centering
\caption{Parameters Definition} \label{ParaDef1}
\begin{tabular}{ |l|l| }
  \hline
 $N$           & The network size\\
 $T$           & The total level of the hierarchy\\
 $N_i^{(l)}$ & The cluster size at level $i$ in cluster $l$\\ 
 $M_i^{(l)}$ & The amount of data transmitted after performing CS at level $i$ in cluster $l$\\
 $C_i$        & The collection of clusters at level $i$\\
 $|C_{i}|$    & The number of cluster at level $i$ in cluster $l$\\
                  & where $|C_{i}| = n^{T-i}$\\
   \hline
\end{tabular}
\end{table}

In order to capture varying sparsity of the data field based on local regions, we also define some new variables.
\begin{itemize}
  \item $K_T$: the whole data field sparsity\\
  \item $K_{i\_T}$: threshold defined as
  $K_{i\_T}= \max_{\substack{l \in C_i}} \{ \frac{N_i^{(l)}}{\log{N_i^{(l)}}} \}$ at level $i$\\
  \item $K_i^{(l)}$: sparsity of the data field in cluster $l$ at level $i$\\    
\end{itemize}

Besides, we also define two types of nodes: CS-enabled nodes and 
CS-disable nodes. In CS-enabled nodes the data collected is large and sparse enough that CS pays off.
On the other hand, in CS-disabled nodes the data collected is small and/or not sparse enough to yield the benefits of CS.  

The salient steps of A-HDACS implemented on the multi-resolution data collection hierarchy are as follows:
\begin{enumerate}

\item At level one, leaf nodes send their single sensed data to their cluster heads without applying CS. 
The cluster head receives the data and performs the conventional transformation to 
obtain the signal representation and its sparsity factor $K_1^{(l)}$. Then
it compares $K_1^{(l)}$ to $\frac{N_1^{(l)}}{\log{N_1^{(l)}}}$.
If $K_1^{(l)} < \frac{N_1^{(l)}}{\log{N_1^{(l)}}}$, it becomes the CS-enabled sensor and takes the CS random 
measurements. The amount of data that need to be transmitted is $M_1^{(l)}=K_1^{(l)} 
\log{N_1^{(l)}}$; otherwise, it disables itself as 
CS-disabled node and transmits $N_1^{(l)}$ data directly to its parent clusters.

\item At level $i$ ($ i \geq 2$), cluster head receives packets from its children nodes. If 
it receives packets with CS random measurements, the CS recovery algorithm will 
be performed firstly to recover all the data.
After cluster head gets all the data from the children nodes, it projects the 
whole data into transformation domain to obtain the signal representation and its 
sparsity factor $K_i^{(l)}$. If $K_i^{(l)} < \frac{N_i^{(l)}}{\log{N_i^{(l)}}}$, 
cluster head turns itself as CS-enabled node and performs the process of taking 
CS random measurements with length $M_i^{(l)}=K_i^{(l)} \log{N_i^{(l)}}$; 
otherwise, it becomes CS-disabled node and send the data directly. 
\label{repeat}

\item Repeat step \ref{repeat} ) until the cluster head at the top level $T$ obtains 
and recovers the whole data.
\end{enumerate}

\section{Analysis of Data Field Sparsity}\label{ana_sparsity}
\newtheorem{prop}{Proposition}\label{prop1}
\begin{prop}
 In HDACS, if $K_T>K_{i\_T}$, all the nodes at the level equal to and below
 $i$ are all CS-disabled nodes. 
\end{prop}
\begin{IEEEproof}
  Define: $f(x)=\frac{x}{\log{x}}$. since $f'(x)=\frac{\log{x}-\frac{1}{\ln{2}}}{(\log{x})^2}>0 \text{ when } 
  x>3$. Therefore, $f(x)$ is a monotonous increasing function when $x>3$.
  \begin{enumerate}
  \item At level $i$, if $K_T>K_{i\_T}$ then $K_T> \frac{N_i^{(l)}}{\log{N_i^{(l)}}}$. In HDACS, CS requires 
	the amount of data to be transmitted $M_i^{(l)} = K_T \log{N_i^{(l)}}$.
        Therefore, $M_i^{(l)} > N_i^{(l)} \text{ for } \forall j \in C_i$. Thus clusters at level $i$ are all CS-disabled nodes.
  
  \item At level $j$ and $j<i$, since $N_j^{(l)} < N_i^{(p)}  \text{ for } \forall l\in C_j \text{ and } \forall p \in C_i$, 
  $K_{i\_T}>K_{j\_T}$. So $K_T>K_{j\_T}>\frac{N_j^{(l)}}{\log{N_j^{(l)}}}$ and $M_j^{(l)} = K_T \log{N_j^{(l)}} > 
  N_j^{(l)}$. Thus the nodes at levels below $i$ are also all CS-disabled nodes.
  \end{enumerate}
\end{IEEEproof}

On the other hand, if $\exists l \in C_i $ s.t. $K_T > K_{i\_T} > \frac{N_i^{(l)}}{\log{N_i^{(l)}}} > K_i^{(l)}$ at level 
$i$. In A-HDACS, since $M_i^{(l)}=K_i^{(l)} \log{N_i^{(l)}} < N_i^{(l)}$,  CS can be utilized. 

Let's define $C_i'$ consisting of all the clusters as CS-disabled nodes at level $i$ in A-HDACS, 
 $\rho_i$ the percentage of CS-disabled clusters at level $i$. In cluster 
 $l$, $\sigma_i^{(l)}$ is defined as the percentage of the CS-disabled children clusters in a CS-disabled 
cluster at level $i$, where $\sigma_i^{(l)} \in \{\frac{1}{n},\frac{2}{n},\cdots, \frac{n}{n} 
\}$. We get $ \rho_i = \frac{|C_i'|}{|C_i|}$ at level $i$; and 
$\rho_{i-1}=\frac{\sum_{l=1}^{|C_i'|} n \sigma_i^{(l)}}{|C_{i-1}|}$ at level $i-1$. 

\begin{prop}\label{prop2}
 If $K_T>K_{i\_T}$, the CS-disabled nodes of A-HDACS at the level equal to and below 
 $i$ are only small percentage of that of HDACS. 
\end{prop}
\begin{IEEEproof}
Let's define $\sigma_i =\frac{1}{|C_i'|}\sum_{l=1}^{|C_i'|} \sigma_i^{(l)} $, 
which shows the average ratio of CS-disabled children clusters to their 
parent clusters. Therefore, we get $\rho_{i-1}=\frac{n |C_i'| \sigma_i}{|C_{i-1}|} = \frac{|C_i'|\sigma_i}{|C_i|} =\rho_i  \sigma_i $. 
Follow the same derivation, $\rho_{i-2}=\rho_i \sigma_i \sigma_{i-1} , 
\rho_{i-3}=\rho_i  \sigma_i \sigma_{i-1}\sigma_{i-2},\cdots, \rho_{1}=\rho_i  \sigma_i 
\sigma_{i-1}\cdots \sigma_{2} $. In summary, the ratio of CS-disabled 
clusters in HDACS at level $i$ and below level $i$ is:
\begin{equation*}
\zeta= \frac{\sum_{j=1}^i |C_j| \rho_j }{\sum_{j=1}^i |C_j|} =
\frac{\sum_{j=1}^i |C_j| \rho_i ( \sigma_i\sigma_{i-1}\cdots \sigma_{j+1} )}{\sum_{j=1}^i |C_j|}
\end{equation*}
Since $\rho_i$ and $\sigma_i$ are equal to or less than 1, $\zeta$ is strictly less than 1. 
Thus, it proves that only $\zeta$ percent of the nodes at the level equal to and below 
 $i$ are CS-disabled nodes for A-HDACS.
\end{IEEEproof}

At the level higher than $i$, i.e. $i < t < T$, the conditions are more diversified and we summarize 
them as follows:
  \begin{enumerate}
    \item If $\frac{N_t^{(l)}}{\log{N_t^{(l)}}} > K_t^{(l)} > K_T $, HDACS and 
    A-HDACS both enable CS. HDACS requires fewer measurements than A-HDACS. 
    But the problem is whether or not HDACS can guarantee the 
    recovery accuracy when a local area has significantly more data variations compared to the global 
    area. \label{cond1}
    \item If $K_t^{(l)} >\frac{N_t^{(l)}}{\log{N_t^{(l)}}} >  K_T $, HDACS 
    enables CS and A-HDACS requires direct data transmission. But it has the 
    the same problem as condition \ref{cond1}). 
    \item If $K_T > \frac{N_t^{(l)}}{\log{N_t^{(l)}}} > K_t^{(l)}$, A-HDACS 
    enables CS but HDACS does not. 
    \item If $\frac{N_t^{(l)}}{\log{N_t^{(l)}}} > K_T > K_t^{(l)} $, both HDACS 
    and A-HDACS enable CS. But HDACS requires more measurements.
    \item The remaining conditions disable CS for both aggregation models. 
  \end{enumerate}  

\begin{figure*}
\centering 
\subfigure[A smooth data field with fluctuations]{\includegraphics[width=2in]{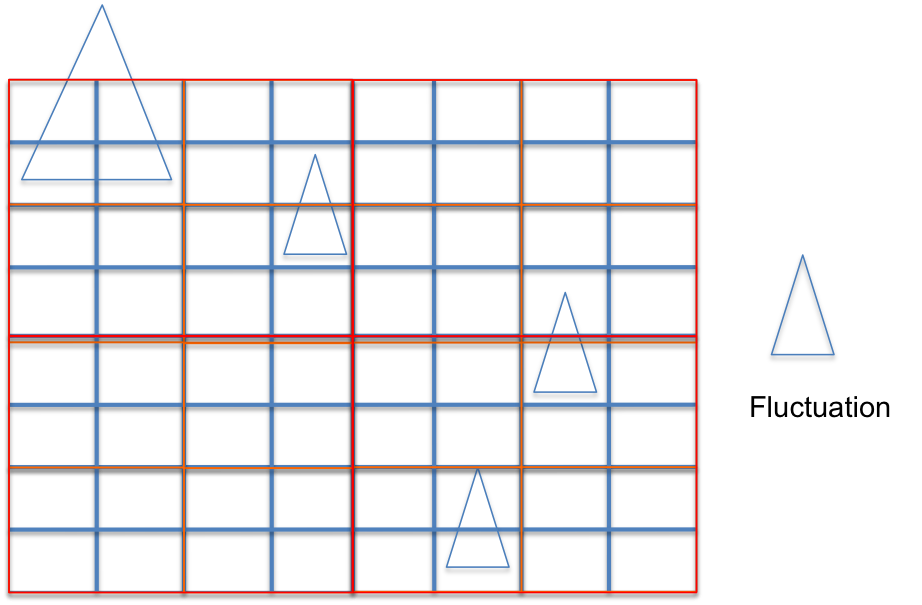}\label{dataField}}
\qquad
\subfigure[HDACS logical tree]{\includegraphics[width=2in]{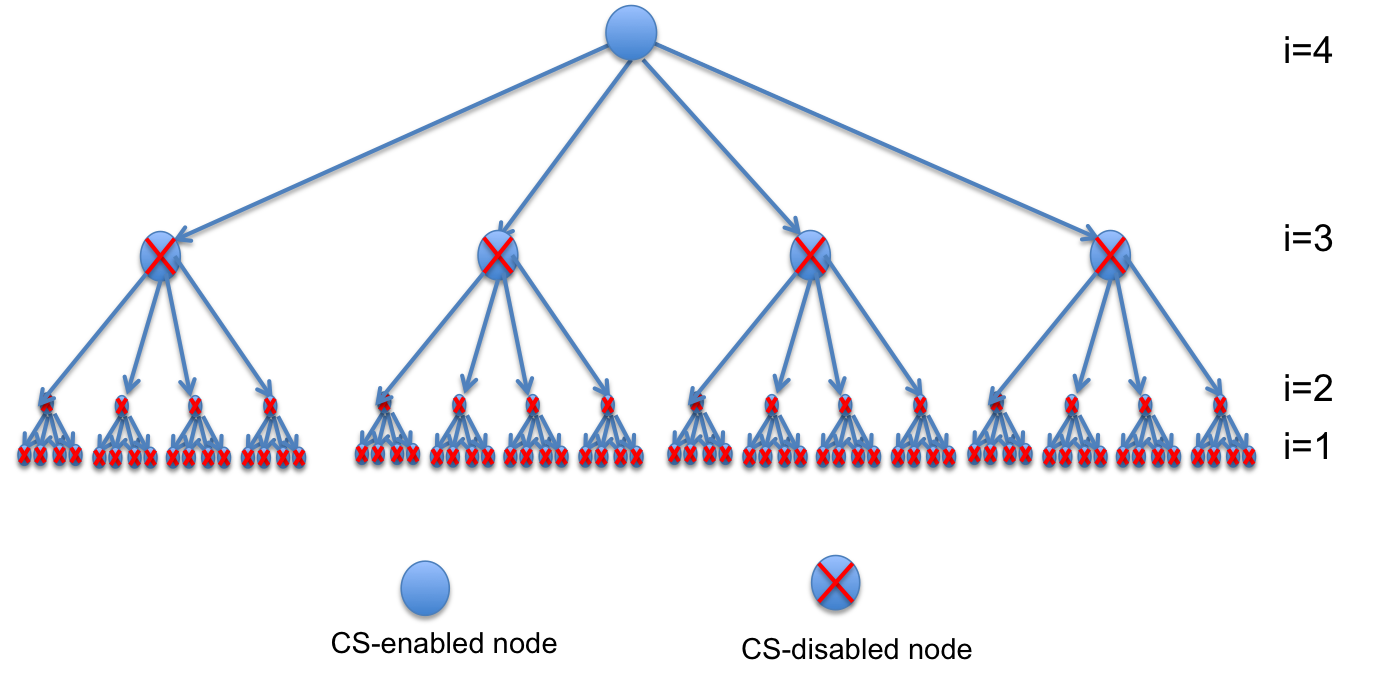}\label{HDACS_logTree}}
\qquad
\subfigure[A-HDACS logical tree]{\includegraphics[width=2in]{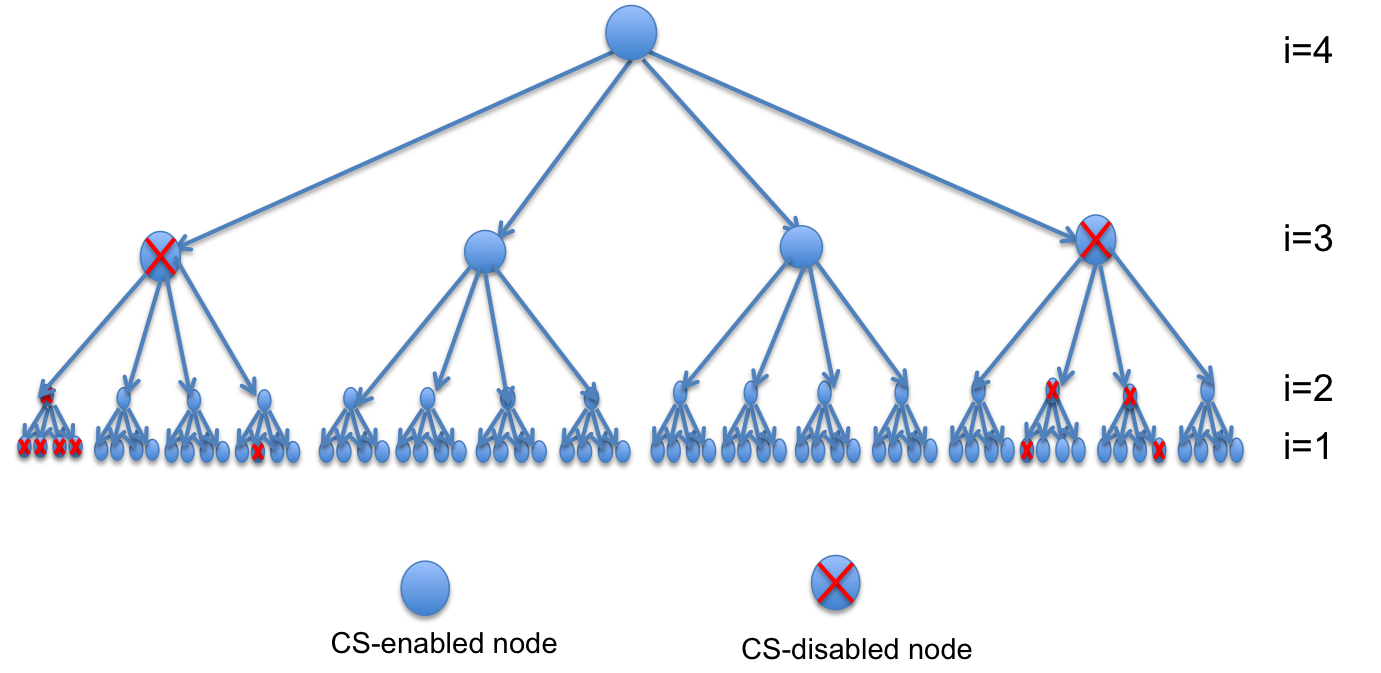}\label{A_HDACS_logTree}}
\caption{An example of a smooth data field with fluctuations and its corresponding logical tree in HDACS and A-HDACS}
\end{figure*}

To better understand this analysis, Fig.\ref{dataField} gives a simple example of a smooth data field 
with a few variations measured by the sensor network in a data aggregation task.
Fig.\ref{HDACS_logTree} and Fig.\ref{A_HDACS_logTree} are its corresponding logical hierarchical trees 
in HDACS and A-HDACS. The local variations in data field lead to 
the large value of global sparsity constant $K_T$ of the data field, and in HDACS it leads to plenty of nodes to be classified 
as CS-disabled 
nodes. However, in the same situation, since in A-HDACS sparsity constants $K_i$s are computed based on local variations in each cluster $i$,
a large fraction of the CS-disabled nodes in HDACS become CS-enabled nodes in A-HDACS. 

\section{Performance Evaluation}
\subsection{Simulation Settings}
\begin{figure*}
\centering
\subfigure[Smooth data field filled with Gaussian bumps]{\includegraphics[width=1.5in]{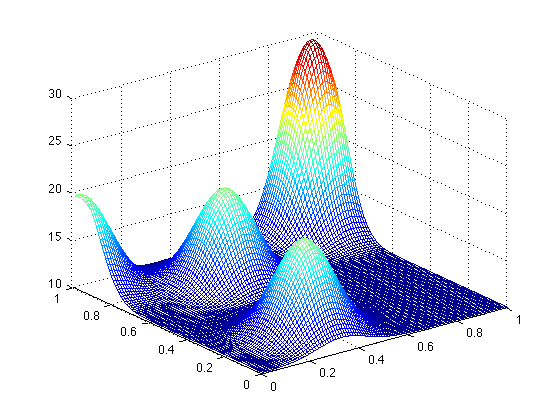}\label{GB}}
\qquad
\subfigure[Piecewise data field]{\includegraphics[width=1.5in]{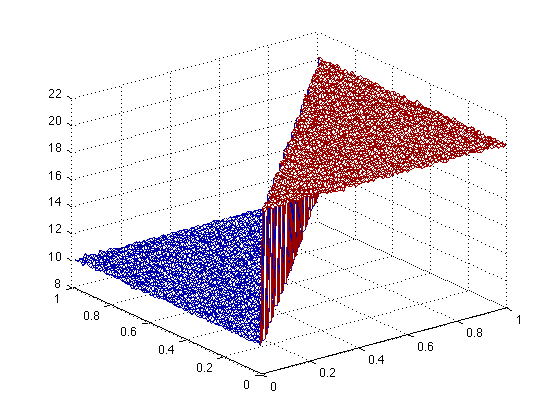}\label{PW}}
\qquad
\subfigure[DCT domain of smooth data field filled with Gaussian bumps]{\includegraphics[width=1.5in]{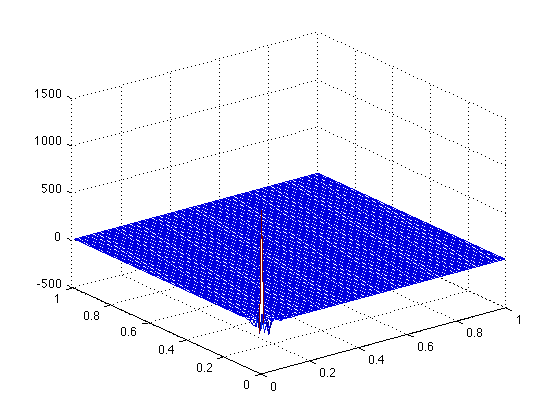}\label{GBDCT}}
\qquad
\subfigure[DCT domain of piecewise data field]{\includegraphics[width=1.5in]{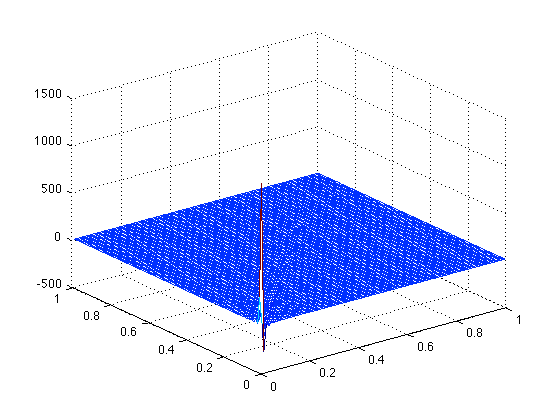}\label{PWDCT}}
\caption{Data Fields and their corresponding DCT Domain}
\end{figure*}

We evaluate the performance of the proposed A-HDACS scheme using SIDnet-SWANS \cite{SIDnet}, a JAVA based sensor network simulation platform. 
In our experiments we have used multiple network sizes, ranging from 300 to 800 sensor nodes, populated in a fixed field size of $4000*4000 m^2$ area.
The average nodes distribution density varies from 
$18.75 / \text{km}^2$ to $50 / \text{km}^2$. Fig. \ref{GB} shows a constant data field filled 
with randomly located Gaussian bumps. It has the maximum height 10 units and decays with 0.01 exponential rate. 
Fig. \ref{PW} depicts a smooth data field with a discontinuity along the line 
$x=y$, where the readings from smooth area are either 10 or 20 plus independent Gaussian noise 
with zero mean and 0.01 variance. 

Besides, Discrete Cosine Transform (DCT) has been used to represent the data field in the transform domain.
DCT is a suboptimal transform for sparse signal representation and 
approaches the ideal optimal transform when the correlation coefficient between 
adjacent data elements approaches unity \cite{DCT_KL}. Fig. \ref{GBDCT} and Fig. \ref{PWDCT} show the results 
when data fields are projected into DCT space. As we can see, most signal energy is captured in a very 
few coefficients, and the magnitudes decay rapidly. Also, note that the DCT signal corresponding to the piecewise data field, 
shown in Fig. \ref{PWDCT}, has less fluctuations than the signal corresponding to the smooth data field with Gaussian bumps, 
shown in Fig. \ref{GBDCT}. 

\subsection{The Nodes Distribution}
\begin{figure*}
\centering
\subfigure[A-HDACS: smooth data field filled with Gaussian bumps]{\includegraphics[width=1.5in]{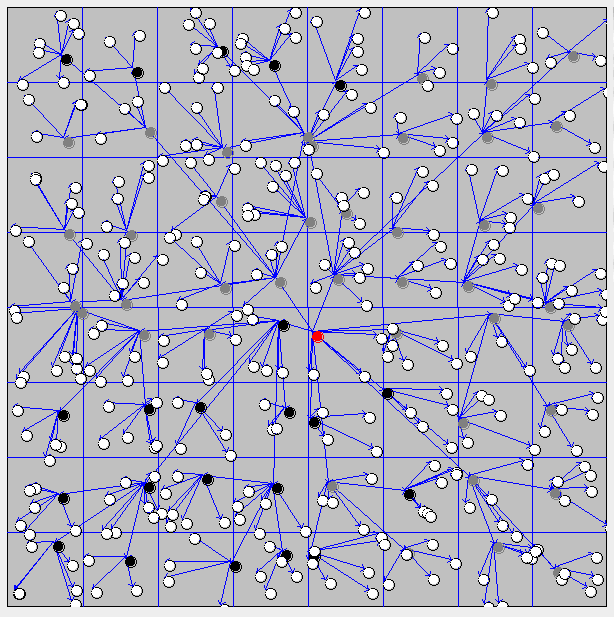}\label{GBSN}}
\qquad
\subfigure[A-HDACS: piecewise data field]{\includegraphics[width=1.5in]{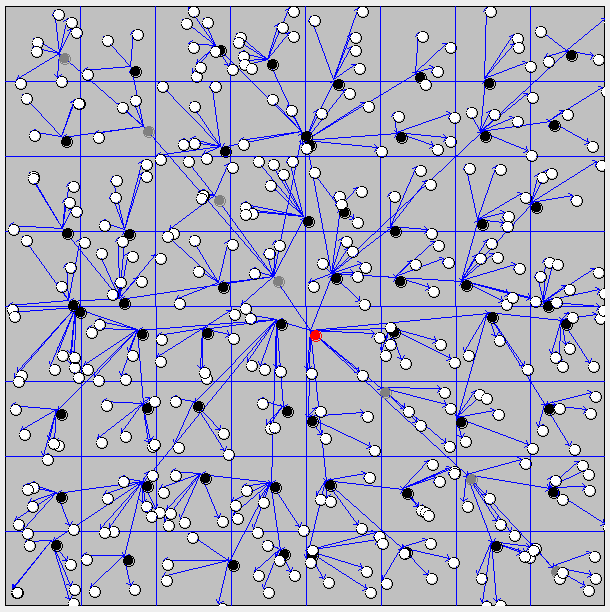}\label{PWSN}}
\qquad
\subfigure[HDACS: smooth data field filled with Gaussian bumps]{\includegraphics[width=1.5in]{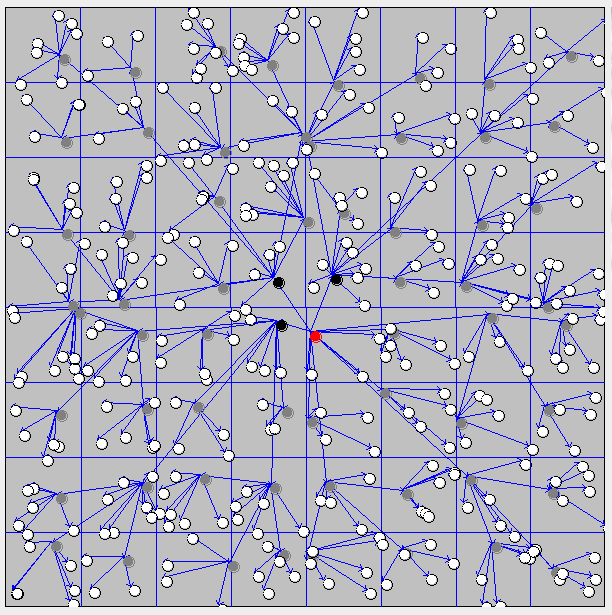}\label{HDACS_GB}}
\qquad
\subfigure[HDACS: piecewise data field]{\includegraphics[width=1.5in]{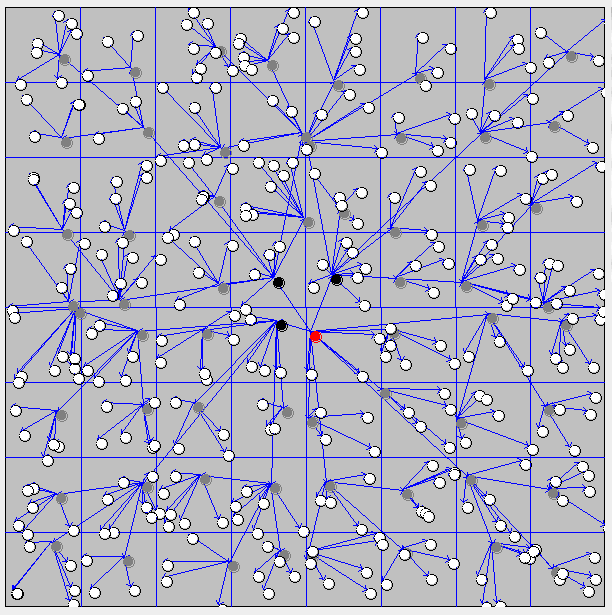}\label{HDACS_PW}}
\caption{The SIDnet simulation results of A-HDACS and HDACS with network size 400: black nodes denote
CS-enabled nodes, gray nodes denote CS-disabled nodes, white nodes are the leaf nodes on level one,
and red node denotes the sink. }
\label{CSdist}
\end{figure*}

Fig. \ref{CSdist} shows the SIDnet simulation results of A-HDACS and HDACS 
for two types of data fields with network size 400, 
where black nodes denote CS-enabled nodes, gray nodes denote that are unable to use CS, and white nodes are the leaf nodes at level one of the aggregation tree. 
As we can see in Fig\ref{GBSN}, 
due to the scattered fluctuations present in the data field with Gaussian bumps  
it is very difficult to obtain sparse signal representation,  
therefore there are only a few CS-enabled nodes. But still for the clusters in 
local smooth data areas A-HDACS is able to utilize CS. 
Fig. \ref{PWSN} shows that 
piecewise data field has a large percent of CS-enabled nodes.
CS-disabled nodes are mainly placed around the discontinuity of the line $x = y$. And the clusters away from this 
line can fully utilize CS.  Fig. \ref{HDACS_GB} and Fig. \ref{HDACS_PW} depict the nodes distribution for 
both data fields using HDACS. The results are identical: almost no CS can be performed at the lower level 
except a few nodes at top levels. It demonstrates the significant improvement of 
CS-enabled nodes in A-HDACS and it is consistent with theoretical analysis in Section \ref{ana_sparsity}.  

\subsection{Data Recovery Results}
Common signals are usually K-compressive -- K entries with significant magnitudes and the other entries rapidly 
decaying to zero. Since K-sparse signal is one requirement of CS, it is 
necessary to perform signal truncation process. In the simulation, we tested different 
signal truncation thresholds so as to get as many CS-enabled nodes as  
possible without compromising too much signal recovery accuracy. Based on the characteristic of DCT signal,
truncation threshold is set up as the percentage of the first dominant magnitude. 

\begin{figure}
  \centering
  \includegraphics[width=2.5in]{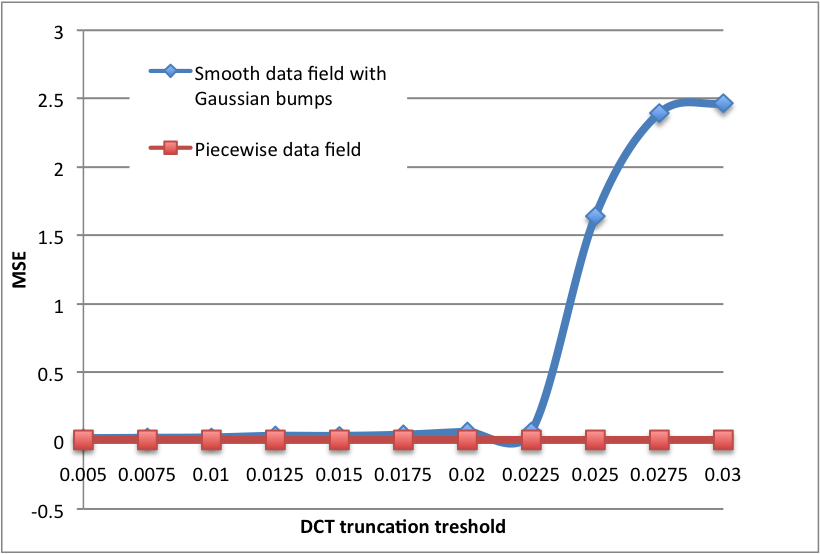}
  \caption{MSE versus DCT truncation threshold with network size 400}
  \label{thresVSmse}
\end{figure}

In the evaluation, Mean Square Error (MSE) of recovered signal in the root node (sink) 
is defined as the average difference between recovered data and actual 
reading values for all the sensors. Fig. \ref{thresVSmse} depicts MSE versus DCT truncation threshold 
for two types of data field with network size 400. 
Since small truncation threshold filters out fewer significant entries than larger thresholds, it 
obtains better MSE. Fig. \ref{thresVSmse} shows that MSE of the smooth data field with Gaussian bumps is below 
0.066 when DCT thresholds are smaller than 0.0225, and it increases 
dramatically when DCT thresholds are large. In the case of the field with Gaussian bumps, fluctuations in the signal
cause increase in the number of DCT coefficients that has significant magnitudes, therefore truncation process is less effective.  
Relatively, piecewise field has more smooth clustering area with only a few significant entries. 
Its MSE is under a negligible value when DCT threshold is in the range of $[0.005, 0.03]$. 
 
\begin{figure}
\centering
\subfigure[Smooth data field filled with Gaussian bumps]{\includegraphics[width=2.5in]{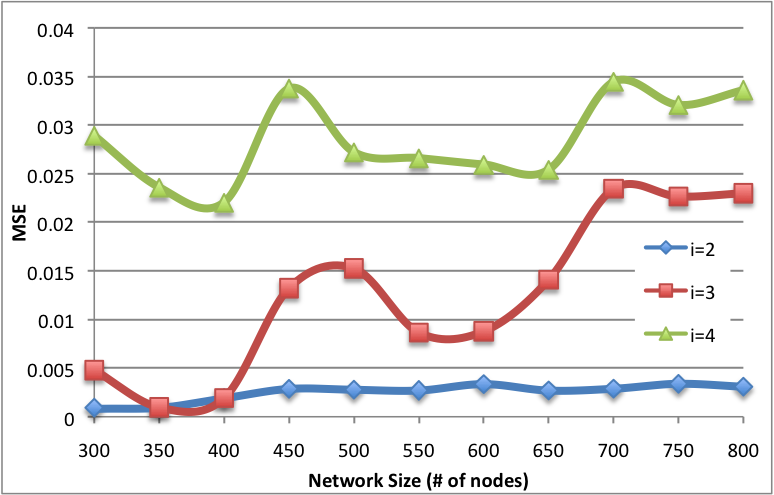}\label{GBMSE}}
\qquad
\subfigure[Piecewise data field]{\includegraphics[width=2.5in]{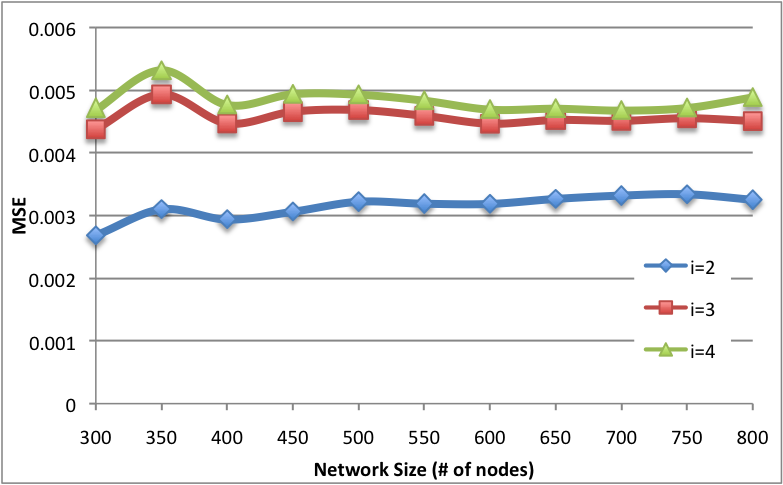}\label{PWMSE}}
\caption{Data recovery mean square error (MSE) results at each level }
\label{MSE}
\end{figure}

In the simulation results presented here onwards, DCT magnitudes bigger than $1\%$ of the first dominant coefficient 
are preserved. Figs. \ref{GBMSE} and \ref{PWMSE} show MSE at each level of the aggregation tree for the two data fields. 
In both cases, MSE results deteriorate with the increase of levels. This is 
because the signal truncation errors propagate in the data aggregation hierarchy. 
In the meanwhile, comparing Fig. \ref{GBMSE} with Fig. \ref{PWMSE}, 
overall piecewise data field has smaller errors than the smooth data field with Gaussian bumps. It is 
due to relatively less fluctuations in the piecewise smooth data field. 

\subsection{Energy Consumption}

\begin{figure}[!t]
\centering
\subfigure[Smooth data field filled with Gaussian bumps ]
{\includegraphics[width=2.5in]{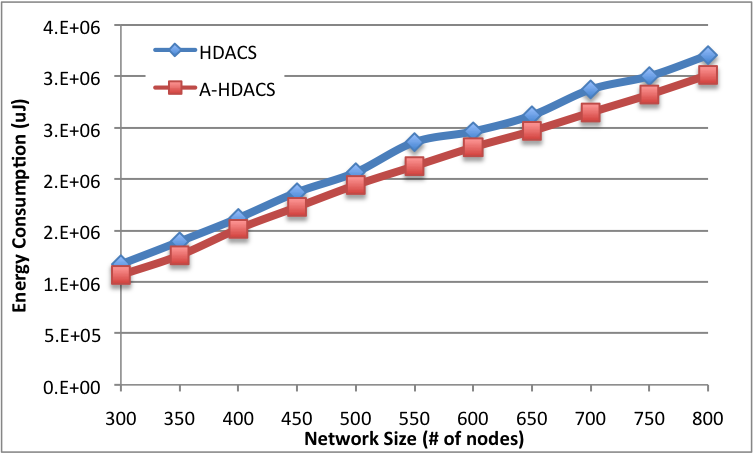}\label{Energy_GB}}
\qquad
\subfigure[Piecewise data field]
{\includegraphics[width=2.5in]{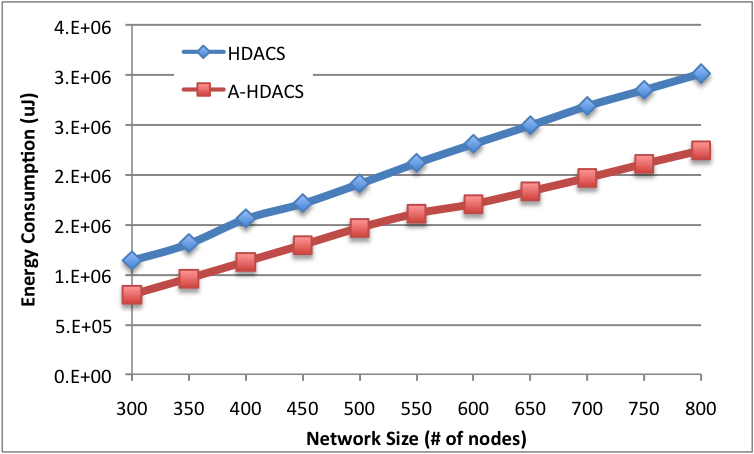}\label{Energy_PW}}
\caption{Total Transmission Energy Cost versus Different Network Sizes}
\label{TotalCost}
\end{figure}

Since communication operations consumes majority of the battery power, we start counting energy consumption 
only when data aggregation begins. Fig. \ref{Energy_GB}  and Fig. \ref{Energy_PW} show energy consumption
versus networks size for two types of data field. A-HDACS consumes only $90.1\% \sim 94.20\% $ 
energy of HDACS in all the network sizes. Although plenty of 
fluctuations in the data field affects A-HDACS to apply CS in a certain degree, 
it still captures the sparsity feature within a few cluster area. But HDACS is 
insensitive to the local area, when the data field slightly change, it loses 
its data compression capability. This advantage is obvious, when it comes to 
the piecewise data field. Fig. \ref{Energy_PW} shows that A-HDACS can save around 
$23.36\% \sim 30.17\%$ battery power, compared to HDACS. The results demonstrate that significant 
energy efficiency can be obtained by the proposed technique.  

\section{Conclusion and Future Work}
In this paper, Adaptive Hierarchical Data Aggregation using Compressive Sensing (A-HDACS) has been 
proposed to perform data aggregation in non-smooth multimodal data fields. Existing CS based data aggregation schemes for WSNs are 
inefficient for such data fields, in terms of energy consumed and amount of data transmitted.
The A-HDACS solution is based on computing sparsity coefficients using 
signal sparsity of data gathered in local clusters.
We analytically prove that A-HDACS enables more clusters to use CS 
compared to conventional HDACS. The simulation evaluated on SINnet-SWANS also demonstrates
the feasibility and robustness of A-HDACS and its significant improvement of 
energy efficiency as well as accurate data recovery results.

In the future work, more factors will be considered to strength A-HDACS. For example, 
in our implementations the cluster size is fixed at each level of the hierarchy. 
It may be possible to further improve communication cost if 
cluster size itself is also set up depending on the local density of the nodes. Besides, temporal correlations in the data may be exploited to further reduce
the amount of data transmitted. Finally, other distributed computing tasks beyond data aggregation, such DFT, DWT, will also be implemented using  
A-HDACS framework, to take advantage of its power-efficient execution. 
\bibliographystyle{IEEEtran}
\bibliography{AdaptiveHDACS_ref}

\end{document}